\documentstyle[aps,prb,epsf,multicol]{revtex}

\begin{document}
\twocolumn[\hsize\textwidth\columnwidth\hsize\csname@twocolumnfalse\endcsname

\title{Effects of doping on thermally excited quasiparticles in the high-$T_c$
superconducting state}
\author{T. Xiang}

\address{Institute of Theoretical Physics, Academia Sinica, P.O.Box 2735, Beijing 
100080, People's Republic of China}

\author{C. Panagopoulos}

\address{Cavendish Laboratory and IRC in Superconductivity, University of 
Cambridge, Madingley Road, Cambridge CB3 OHE, United Kingdom}

\date{\today}

\maketitle

\begin{abstract}
The physical properties of low energy superconducting quasiparticles in
high- $T_c$ superconductors are examined using magnetic penetration depth
and specific heat experimental data. We find that the low energy density of
states of quasiparticles of La$_{2-x}$Sr$_x$CuO$_4$ scales with $\left(
x-x_c\right)/T_c$ to the leading order approximation, where $x_c $ is the
critical doping concentration below which $T_c=0$. The linear temperature
term of the superfluid density is renormalized by quasiparticle interactions
and the renormalization factor times the Fermi velocity is found to be
doping independent. 
\end{abstract}


]

In high-$T_c$ superconductors (HTS), the low energy density of states (DOS)
is linear due to the $d_{x^2-y^2}$-wave pairing symmetry\cite{Tsuei}. This
linear DOS leads to a linear in-plane superfluid density (which is
proportional to the inverse square of the magnetic penetration depth $
\lambda _{ab}^{-2}$)\cite{Hardy,PanaHg} and a quadratic electronic specific
heat\cite{LoramYB} at low temperatures. The experimental observations of
these power-law temperature dependences provided some of the early evidence
for the unconventional pairing symmetry and lent support to the Fermi liquid
description of the high-$T_c$ superconducting state. Exploring the physical
properties of the low energy quasiparticle excitations is of fundamental
importance for understanding the high-$T_c$ mechanism. A central issue which
is currently under debate is the nature of quasiparticle interactions and
their effect on the physics of the superconducting state\cite
{Lee,Millis98,Pana98}. Recently Mesot et al\cite{MesotVF} calculated the
slope of the superfluid stiffness using parameters obtained from angular
resolved photoemission (ARPES) measurements for Bi$_2$Sr$_2$CaCu$_2$O$_{8+x}$
(BSCCO). They found that the renormalization factor to the superfluid
stiffness is doping dependent and about a factor of 2 to 3 smaller than 
that for non-interacting quasiparticle systems.

In this paper we provide further evidence for the existence of strong
quasiparticle interaction in the superconducting state and present a
detailed analysis for the low energy DOS and other fundamental parameters of
HTS. We calculate the doping dependence of these parameters using high
quality penetration depth and specific heat data and discuss an interesting
correlation between an energy scale derived from the low temperature
superfluid response and the normal state pseudogap.

Let us first consider the low energy DOS of high-$T_c$ quasiparticles $
N(\omega )$. If we denote the linear coefficient of $N(\omega )$ by $\eta $,
then at low energies $N(\omega )$ is given by 
\begin{equation}
N(\omega )\approx \eta \omega .
\end{equation}
Since the low temperature penetration depth and specific heat are governed
by thermally excited quasiparticles, it can be shown that the slopes of $
\lambda ^{-2}$ and $\gamma $ at zero temperature are given by 
\begin{eqnarray}
{\frac{d\lambda _{ab}^{-2}}{dT}}|_{T\rightarrow 0} &=&(4\pi \ln 2)\eta
k_B\left( \frac{e\beta v_F}c\right) ^2,  \label{lambda} \\
{\frac{d\gamma }{dT}}|_{T\rightarrow 0} &=&5.4\eta k_B^3,  \label{gamma}
\end{eqnarray}
where $\gamma =C_v/T$ is the specific heat coefficient and $v_F$ is the
Fermi velocity. $\beta $ is a renormalization factor to the paramagnetic
term in the superfluid density due to quasiparticle interactions or vertex
corrections\cite{Millis98}. For non-interacting quasiparticles $\beta =1$,
and Eq. (\ref{lambda}) is the standard BCS mean-field result\cite
{Hirschfeld,Xiang}. The above equations show that from experimental data of $
\lambda _{ab}$ and $C_v$, we can determine the values of two important
parameters: the linear coefficient of DOS $\eta $, and the product of the
renormalization constant and Fermi velocity $\beta v_F$.

Figure 1 shows the experimental result of $T_cd\gamma /dT$ at $T=0K$ as a
function of doping for La$_{2-x}$Sr$_x$CuO$_4$ (LSCO)\cite{Loram,Momono}.
Note here we plot the product $T_cd\gamma /dT$ rather than $d\gamma /dT$
since in an ideal BCS superconductor with $d$-wave pairing, $T_cd\gamma
/dT(0K)$ is proportional to the charge concentration. We find that $
T_cd\gamma /dT(0K)$ increases monotonically with doping and the slope is
smaller in the underdoped regime than in the overdoped one. The value of $
\eta $ can be obtained from the data shown in Fig. 1. In the whole doping
regime, we find that $\eta $ can be approximately fitted by 
\begin{equation}
\eta \approx {\frac{\eta _1(x-x_c)+\eta _2(x-x_c)^2}{T_c},}
\end{equation}
where $x_c$ $\sim $ $0.058$ is approximately equal to the critical 
doping concentration below which $T_c=0$, 
$\eta _1=13$ $mJ/(K^2k_B^3mol)$ and $\eta _2=87$ $mJ/(K^2k_B^3mol)$. For
other high-$T_c$ materials, the low temperature data for $\gamma $ generally
contain a significant contribution from magnetic impurities and it is
difficult, especially in the underdoped regime, to determine accurately the
value of $d\gamma /dT(0K)$. However, from the data available we find that,
within experimental errors, the doping dependence of $\eta $ is similar to
that shown in Fig. 1.

\begin{figure}
\leavevmode\epsfxsize=8cm
\epsfbox{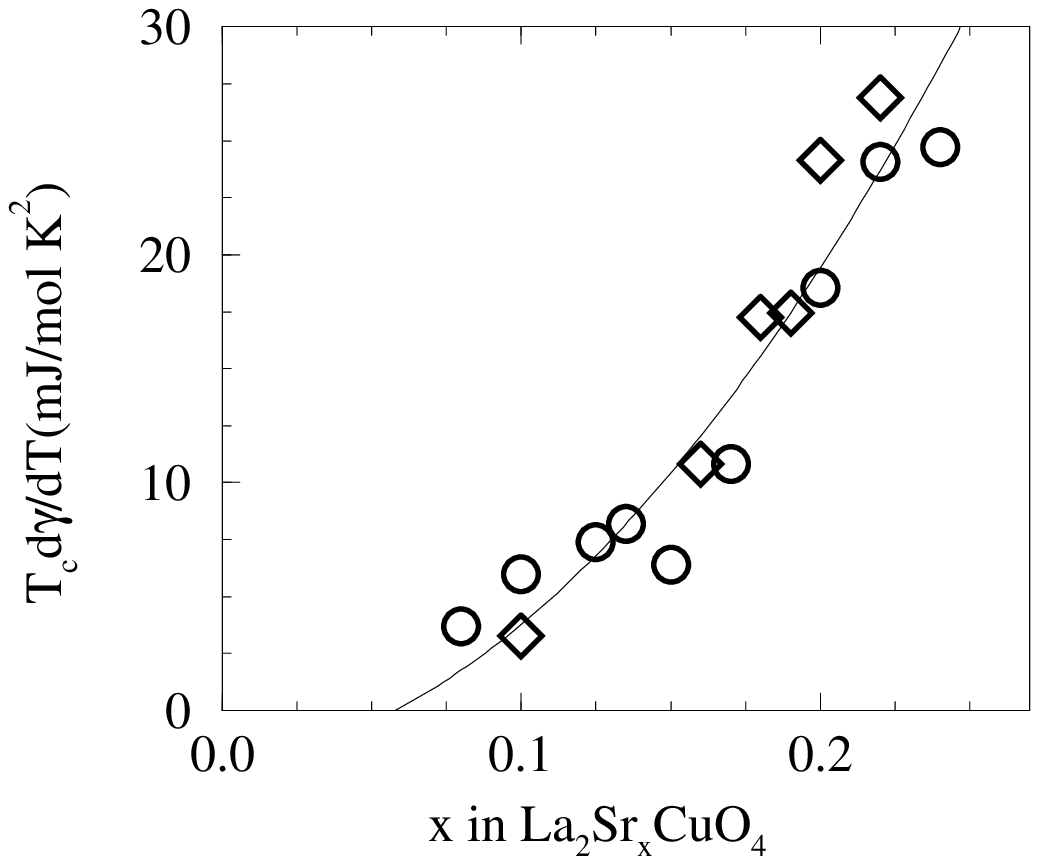}
\caption{$T_cd\gamma /dT$ as a function of doping for ${\rm 
La_{2-x}Sr_xCuO_4 }$. (Circles are from Ref. \cite{Loram} and diamonds are from
Ref. \cite{Momono}.) The solid curve is a least square fit to the experimental data with $\eta$ given by Eq. (4). }
\label{fig1}
\end{figure}

Figure 2 shows the extrapolated experimental results of $T_cd\lambda
_{ab}^{-2}/dT$ at $0K$ as a function of doping, $x$, for LSCO and YBa$_2$Cu$
_3$O$_{6+x}$ (YBCO)\cite{PanaLSC,PanaPC,PanaYBC,Bonn}. The magnetic
penetration depth data for LSCO and the specific heat data of Loram {\it et
al }shown in Fig. 1 were obtained from samples of the same batch. The
similar doping dependence in $T_cd\lambda _{ab}^{-2}/dT$ and $T_cd\gamma /dT$
indicates that it is indeed the thermally excited quasiparticles which are
responsible for the low temperature thermodynamic response of HTS. Two sets
of penetration depth data are shown for YBCO. One set comes from the $ac$
-susceptibility measurements of grain-aligned YBCO\cite{PanaPC,PanaYBC} and
the other from microwave measurements of detwinned YBCO single crystals \cite
{Bonn}. The $ac$-susceptibility technique measures the effective in-plane
penetration depth, $\lambda _{ab}$, whereas the microwave experiment
measures the penetration depth along two principal axes, $\lambda _a$ and $
\lambda _b$. In order to compare these two sets of data, we have converted
the single crystal data to the effective in-plane penetration depth by
assuming $\lambda _{ab}^2\approx \lambda _a\lambda _b$ \cite
{Bradford,Pana124}. The agreement between the two YBCO data sets is striking
considering the difference in the type of samples and measurement techniques
used by the two groups. For both LSCO and YBCO we find that within
experimental errors $T_cd\lambda _{ab}^{-2}/dT$ increases almost linearly
with doping.

\begin{figure}
\leavevmode\epsfxsize=8cm
\epsfbox{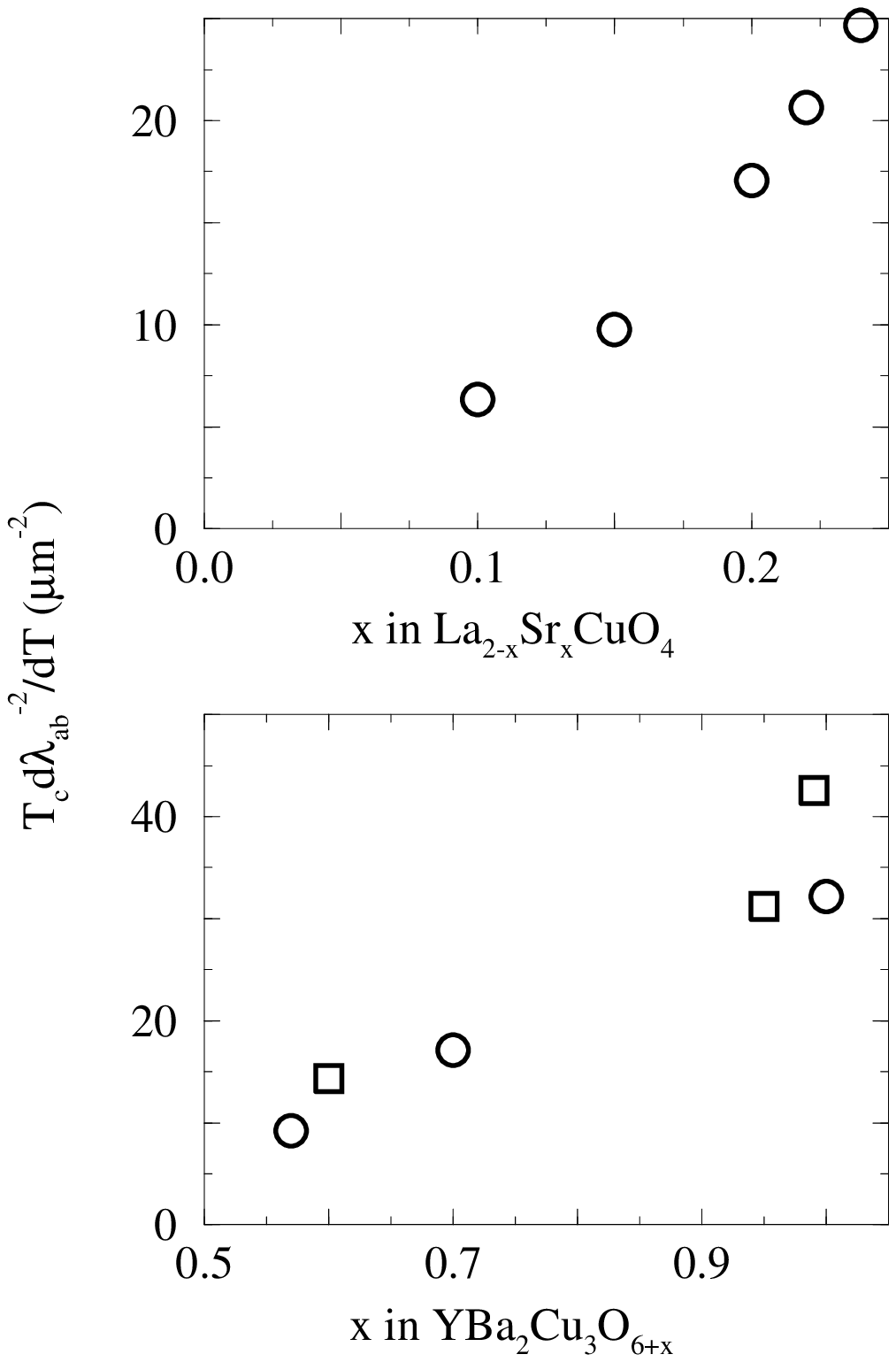}
\caption{$T_cd\lambda _{ab}^{-2}/dT$ as a function of doping for ${\rm 
La_{2-x}Sr_xCuO_4}$ (upper panel)\protect\cite{PanaLSC} and ${\rm 
YBa_2Cu_3O_{6+x}}$ (lower panel). In the lower panel, circles are from Ref. 
\protect\cite{PanaPC,PanaYBC}, and squares are from Ref. \protect\cite{Bonn}
. }
\label{fig2}
\end{figure}

Using the penetration depth and specific heat data for LSCO we have
estimated $\beta v_F$ from the ratio of $d\lambda ^{-2}/dT$ and $d\gamma /dT$
. As shown in Fig. 3, $\beta v_F$ is almost doping independent and of the
order $5$ $\sim $ $6\times 10^6cm/\sec $. If we assume that the Fermi
velocity of LSCO is equal to the Fermi velocity of BSCCO, i.e. $
v_F=2.5\times 10^7cm/\sec $, \cite{DingVF,MarshallVF,MesotVF,Lee}, Fig. 3
suggests that $\beta $ is approximately equal to $0.2$, a value much smaller
than that for non-interacting quasiparticles where $\beta =1$. Obviously
this estimate for $\beta $ is crude since the Fermi velocity for LSCO may
not be the same as that for BSCCO. Nevertheless, it indicates that the
quasiparticle interaction is quite strong, in qualitative agreement with the
analysis of Mesot $et.$ $al$ for BSCCO\cite{MesotVF}. Furthermore, if $v_F$
in LSCO is doping independent as in BSCCO\cite{MesotVF} the result in Fig. 3
suggests that $\beta $ is also doping independent, which disagrees with the
conclusions of Mesot $et$. $al.$ for BSCCO\cite{MesotVF}. We see no apparent
physical reason why the doping dependence of $\beta v_F$ in these two high-$
T_c$ systems should be different. It is likely that the difference is just
due to the experimental uncertainty. To clarify this problem, more
measurements of the doping dependence of the Fermi velocity and low energy
DOS of quasiparticles for both systems are required.

\begin{figure}
\leavevmode\epsfxsize=8cm
\epsfbox{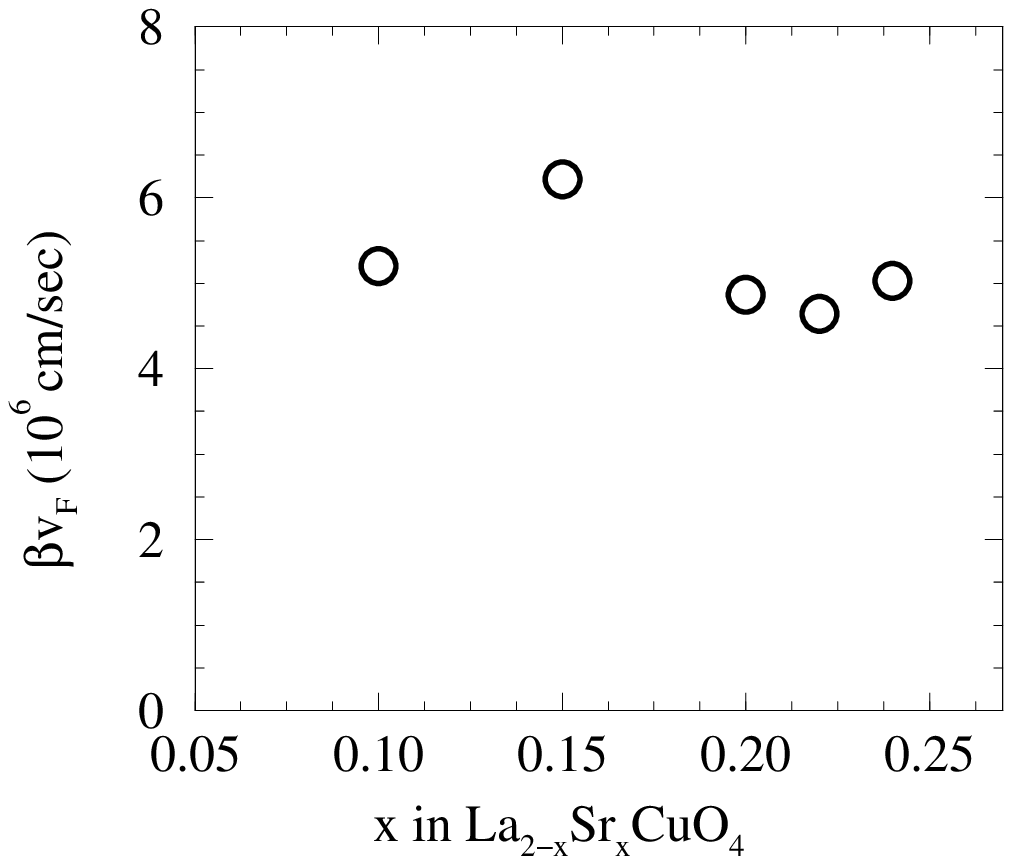}
\caption{$\beta v_F$ estimated from penetration depth and electronic
specific heat measurements for ${\rm La_{2-x}Sr_xCuO_4}$. }
\label{fig3}
\end{figure}

Since $\beta v_F$ for LSCO is approximately doping independent, the above
results indicate that in the underdoped regime 
$\lambda _{ab}^{-2}(0)-\lambda
_{ab}^{-2}(T)\sim (x-x_c)T/T_c$. Experimentally, it was found that $\lambda
_{ab}^{-2}(0)$ is approximately proportional to $T_c$ \cite{Uemura}, this
therefore gives 
\begin{equation}
\frac{\lambda _{ab}^{-2}(T)}{\lambda _{ab}^{-2}(0)}\approx 1-\alpha \frac{
x-x_c}{T_c^2}T,
\end{equation}
where $\alpha $ is a doping independent constant. This equation holds only
at low temperatures and low doping. However, if we extrapolate $\lambda
_{ab}^{-2}(T)$ to high temperatures using this equation, we find that $
\lambda _{ab}^{-2}(T)$ becomes zero at a temperature $T_0\sim T_c^2/(x-x_c)$. 
It is interesting to note that $T_0$ has approximately the
same doping dependence and order of magnitude as the temperature scale $
T^{*} $ of the normal state gap obtained from tunnelling, ARPES, and other
measurements \cite{Timusk}. If we interpret $T_0$ as the energy
scale at which Cooper pairs begin to form and the difference between $T_0$
at $T_c$ is due to the pair phase fluctuations this result seems to be consistent with
the widely discussed phase fluctuation picture\cite{Emery}. 
However, as both the XY order parameter fluctuations and 
the gauge fluctuations are important in the 
critical phase transition regime, this interpretation is not unique.  
Nevertheless, it suggests that from low temperature measurements of the 
superconducting state, one can also obtain information on phase fluctuations in the normal state. 

In conclusion, using low temperature experimental data of the magnetic
penetration depth and electronic specific heat of HTS, we have estimated the
effect of doping on the low energy DOS of quasiparticles and $\beta v_F$.
Both $T_cd\gamma /dT$ and $T_cd\lambda _{ab}^{-2}/dT$ are found to increase
with increasing doping whereas $\beta v_F$ is doping independent. The low
temperature superfluid density $\lambda _{ab}^{-2}(T)$ extrapolates to zero
at a temperature $T_0$ which has approximately the same doping dependence
and order of magnitude as the onset temperature of the normal state gap.

We would like to thank J. W. Loram for supplying the specific heat data. CP
thanks Trinity College, Cambridge for financial support.
T.X. was partially supported by the Natural Science Fundation of China.


\end{document}